\documentclass[aps,prd,onecolumn,groupedaddress,showpacs,nofootinbib,amssymb]{revtex4}
\usepackage[dvips]{graphicx}
\usepackage{amssymb}
\usepackage{amsmath}
\usepackage{graphicx,,color}
\usepackage{amsfonts}
\usepackage{bm}
\usepackage{cancel}
\usepackage{comment}

\allowdisplaybreaks[4]

\begin{document}

\tolerance=5000

\newcommand\be{\begin{equation}}
\newcommand\ee{\end{equation}}
\newcommand\nn{\nonumber \\}
\newcommand\e{\mathrm{e}}

\title{Cosmological Bound from the Neutron Star Merger GW170817 \\ 
in scalar-tensor and $F(R)$ gravity theories}

\author{Shin'ichi Nojiri$^{1,2,3}$, Sergei D. Odintsov$^{4,5,6,7,8}$}

\affiliation{
$^1$ Department of Physics,
Nagoya University, Nagoya 464-8602, Japan\\ $^2
$ Kobayashi-Maskawa Institute for the Origin of Particles and the Universe, Nagoya University, 
Nagoya 464-8602, Japan\\ 
$^3$ KEK Theory Center, High Energy Accelerator Research Organization (KEK), Oho 1-1, 
Tsukuba, Ibaraki 305-0801, Japan\\ 
$^4$ICREA, Passeig Luis Companys, 23, 08010 Barcelona, Spain, \\ 
$^5$ Inst. Space Sciences (ICE-CSIC), C. Can Magrans s/n, 08193 
Barcelona, Spain, \\
$^6$ Inst. Space Sciences of Catalonia (IEEC), 08034 Barcelona, Spain, \\
$^7$ National Research Tomsk State University, 634050 Tomsk, Russia, \\
$^8$ Tomsk State Pedagogical University, 634061 Tomsk, Russia
}

\date{\today}
\begin{abstract}
We consider the evolution of cosmological gravitational waves in 
scalar-tensor theory and $F(R)$ gravity theory as typical models of the modified gravity.
Although the propagation speed is not changed from the speed of light, the propagation phase changes 
when we compare the propagation in these modified gravity theories with the propagation in the 
$\Lambda$CDM model.
The phase change might be detected in future observations. 
\end{abstract}


\maketitle

\section{Introduction}

The early-time inflation  and late-time acceleration  are eventually most important problems in physics of 
this century.
In order to solve these problems, number of models, including the theories modifying the Einstein gravity, 
have been proposed and investigated (for the review, see 
\cite{Nojiri:2017ncd,Capozziello:2011et,Nojiri:2010wj,Bamba:2012cp,
Capozziello:2010zz}).
The fundamental problem is to identify the model which most naturally describes the universe.
Although we have obtained some constraint from the cosmology \cite{Ade:2015rim}, the recent 
discovery of the gravitational wave \cite{Abbott:2016blz,Abbott:2016nmj}, may also indicate the 
possibility to select the most realistic theory or give some constraints on the model by using the 
observational data \cite{DeLaurentis:2016jfs,Capozziello:2008rq,Capozziello:2008fn,
Bellucci:2008jt,Bogdanos:2009tn}.
Especially, in \cite{Capozziello:2017vdi}, it has been shown that the amplification of the amplitude of 
gravitational wave changes in the scalar-tensor theory and $F(R)$ gravity if compared with the Einstein 
gravity (for the modification in the brane world or the domain wall universe, see 
\cite{Higuchi:2014bya}).

In fact,
the recent observation of GW170817 \cite{TheLIGOScientific:2017qsa}, where the gravitational wave 
was generated by the neutron stars merger, may put some constraints to different cosmological theories.
In the observation of GW170817, not only the gravitational wave but also the signal of the gamma ray 
has been detected simultaneously. This gives the strong constraint on the ratio for the speeds of  the 
gravitational wave and the light, 
\begin{equation} 
\label{GWp9} \left| 
\frac{c_\mathrm{GW}^2}{c^2} - 1 \right| < 6 \times 10^{-15}\, .
\end{equation}
Here $c$ is the speed of the light and $c_\mathrm{GW}$  is the propagating speed of the gravitational wave.
Eq.~(\ref{GWp9})  constrains  the mass of the graviton and the parameters in the scalar-tensor theory 
of Horndeski or Galileon type model 
\cite{Lombriser:2015sxa,Bettoni:2016mij,Creminelli:2017sry,Sakstein:2017xjx,Baker:2017hug}.
In case of the covariant Galileon model \cite{Deffayet:2009wt}, whose Lagrangian density is given by 
\begin{align} 
\label{GWp12} 
\mathcal{L} =& X + G_4(X)R + G_{4,X} \left( \left( \nabla^2 \phi \right)^2
 - \nabla_\mu \nabla_\nu \phi \nabla^\mu \nabla^\nu \phi \right) \, , \quad 
X = - \frac{1}{2} \partial_\mu \phi \partial^\mu \phi \, , \nn
G_4(X) =& \frac{M_\mathrm{Pl}^2}{2} + \frac{2c_0}{M_\mathrm{Pl}} \phi
+ \frac{2C_4}{\Lambda_4^6} X^2 \, ,
\end{align}
(for the details of the notation, see \cite{Deffayet:2009wt}), the terms including $c_4$, where the 
derivative term of the scalar field couples with the curvature, induce the modification of the effective 
metric for the gravitational wave, 
\begin{equation} 
\label{GWp13} 
g_{\mu\nu} \to g_{\mu\nu} + C \partial_\mu \phi \partial_\nu \phi \, .
\end{equation}
It gives the variation of the speed of the propagation of the gravitational wave, 
\begin{equation} 
\label{GWp14} 
\left| \frac{c_\mathrm{GW}^2}{c^2} - 1 \right| = \left| \frac{4c_4 x^2}{1 - 3 c_4 x^2} 
\right|\, , \quad x= \frac{\dot\phi}{HM_\mathrm{Pl}} \, .
\end{equation}
Therefore Eq.~(\ref{GWp9}) gives a strong constraint on the model.
It is the purpose of our work to investigate whether even if the derivative term of the scalar field does not 
couple with the curvature, there appears the modification in the propagation of the gravitational wave.
Note that in case of Moffat's modified gravity (MOG), that is, Scalar-Tensor-Vector-Gravity (STVG) 
\cite{Moffat:2005si} (see also \cite{Moffat:2013sja,Moffat:2013uaa}), which is also a typical modified 
gravity model, the consistency with the  GW170817 event \cite{TheLIGOScientific:2017qsa} has been 
confirmed in \cite{Green:2017qcv}.

In the next section, we give a general setup for the propagation of the gravitational wave.
In Section \ref{SIII}, we consider the scalar-tensor theory. We show that the the propagation phase of 
the gravitational wave changes in the background although the propagation speed is the same light 
speed.
In Section \ref{SIV}, we consider the propagation of the gravitational wave and we show that the 
propagation is different from that in the scalar-tensor theory.

\section{Setup \label{SII}}

Let us consider the equation for the propagation of gravitational wave in the expanding universe.
The gravitational wave is given by the perturbation from the background geometry, 
\begin{equation} 
\label{H1} 
g_{\mu\nu} \to g_{\mu\nu} + h_{\mu\nu}\, , 
\end{equation} 
where $|h_{\mu\nu}|\ll 1$ is the perturbation with respect to a given background $g_{\mu\nu}$.
It is straightforward to obtain  the perturbed Ricci tensor and scalar, 
\begin{align} 
\label{H2} 
\delta R_{\mu\nu} = & \frac{1}{2} \left[ \nabla_\mu \nabla^\rho h_{\nu\rho}
+ \nabla_\nu \nabla^\rho h_{\mu\rho} - \nabla^2 h_{\mu\nu}
 - \nabla_\mu \nabla_\nu \left( g^{\rho\lambda} h_{\rho\lambda} \right)
 - 2 R^{\lambda\ \rho}_{\ \nu\ \mu} h_{\lambda\rho}
+ R^\rho_{\ \mu} h_{\mu\nu} + R^\rho_{\ \nu} h_{\rho\mu} \right] \, ,\\
\label{H3}
\delta R = & - h_{\mu\nu} R^{\mu\nu} + \nabla^\mu \nabla^\nu h_{\mu\nu}
 - \nabla^2 \left( g^{\mu\nu} h_{\mu\nu}\right)\, .
\end{align}
As long as we consider the propagation of the gravitational wave, we need not to consider the 
perturbation  of  scalar modes because the spin two field, i.e. the graviton corresponding to the 
gravitational wave, does not mix with the scalar field (spin zero) field.
By imposing the gauge condition
\begin{equation}
\label{H4}
\nabla^\mu h_{\mu\nu} = g^{\mu\nu} h_{\mu\nu} = 0\, , 
\end{equation} 
the Einstein field equations 
\begin{equation} 
\label{einstein} 
R_{\mu\nu} - \frac{1}{2} g_{\mu\nu} R = \kappa^2 T_{\mu\nu}\, , 
\end{equation} 
take the perturbed form as follows, 
\begin{equation} 
\label{H6} 
\frac{1}{2} \left[ - \nabla^2 h_{\mu\nu}
 - 2 R^{\lambda\ \rho}_{\ \nu\ \mu} h_{\lambda\rho}
+ R^\rho_{\ \mu} h_{\rho\nu} + R^\rho_{\ \nu} h_{\rho\mu}
  - h_{\mu\nu} R + g_{\mu\nu} R^{\rho\lambda} h_{\rho\lambda} \right] 
= \kappa^2 \delta T_{\mu\nu}\, .
\end{equation}
The equation (\ref{H6}) indicates that the  energy-momentum tensor perturbations affect the 
propagation of gravitational wave.

\section{Scalar-Tensor Theory \label{SIII}}

In order to specify the explicit form of $\delta T_{\mu\nu}$, we consider the scalar theory whose action 
is given by 
\begin{equation} 
\label{H7} 
S_\phi = \int d^4 x \sqrt{-g} \mathcal{L}_\phi\, , \quad \mathcal{L}_\phi 
=  - \frac{1}{2} \omega(\phi) \partial_\mu \phi \partial^\mu \phi - V(\phi) \, .
\end{equation}
One finds
\begin{equation}
\label{H8}
T_{\mu\nu} = - \omega(\phi) \partial_\mu \phi \partial_\nu \phi
+ g_{\mu\nu} \mathcal{L}_\phi\, ,
\end{equation}
and therefore
\begin{equation}
\label{H9}
\delta T_{\mu\nu} = h_{\mu\nu} \mathcal{L}_\phi
+ \frac{1}{2} g_{\mu\nu} \omega(\phi) \partial^\rho \phi
\partial^\lambda \phi h_{\rho\lambda}\, , 
\end{equation} 
up to first order in perturbation.
We are interested  in the evolution of tensor gravitational wave so we can consider only the spatial 
component of $h_{\mu\nu}$, that is, $h_{ij}$.

Assuming a FRW spatially flat metric
\begin{equation}
\label{FRWmetric}
ds^2 = - dt^2 + a(t)^2 \sum_{i=1,2,3} \left( dx^i \right)^2 \, , 
\end{equation} 
and  $\phi=t$ in  (\ref{H9}), the FRW equations have the following form, 
\begin{equation} 
\label{LB1} 
\frac{3}{\kappa^2} H^2 = \frac{\omega}{2} + V\, ,\quad
 - \frac{1}{\kappa^2} \left( 2\ddot H + 3 H^2 \right) = \frac{\omega}{2} - V\, , 
\end{equation} 
and one gets 
\begin{equation} 
\label{LB2} 
\omega = - \frac{2}{\kappa^2} \dot H \, , \quad V = \frac{1}{\kappa^2} 
\left( \dot H + 3 H^2 \right) \, , 
\end{equation} 
for the kinetic term and the scalar field potential expressed as functions of the scale factor $a(t)$ and its 
derivatives.
By using (\ref{H6}), (\ref{H9}), (\ref{LB2}), and $\phi=t$, we find the evolution equation of 
gravitational wave:
\begin{equation}
\label{LB3}
0=\left(2\dot H + 6 H^2 + H \partial_t-\partial_t^2
+\frac{\bigtriangleup}{a^2}\right)h_{ij}\, ,
\end{equation}
which clearly depends on the cosmological background.

We may compare the propagation of the gravitational wave with that of the light or photon.
The equation for the vector field corresponding to the photon is given by 
\begin{equation} 
\label{phtn1}
0 = \nabla^\mu F_{\mu}^{\ \nu}
= \frac{1}{\sqrt{-g}} \partial_\mu \left( \sqrt{-g} g^{\mu\rho} g^{\nu\sigma} F_{\rho\sigma}
\right) = \nabla^2 A^\nu - \nabla^\nu \nabla^\mu A_\mu + R^{\mu\nu} A_\mu \, , 
\quad F_{\mu\nu} = \partial_\mu A_\nu - \partial_\nu A_\mu \, .
\end{equation}
In the FRW universe, Eq.~(\ref{phtn1}) can be rewritten as follows, 
\begin{equation} 
\label{phtn2}
0 = \sum_{i=1,2,3} \partial_i \left( \partial_i A_t - \partial_t A_i \right) \, , \quad
0 = \left(\partial_t + H \right) \left( \partial_i A_t - \partial_t A_i \right)
+ a^{-2} \left( \bigtriangleup A_i - \partial_i \sum_{j=1,2,3}\partial_j 
+ A_j \right)\, .
\end{equation}
We may choose the Landau gauge
\begin{equation}
\label{phtn3}
0 = \nabla^\mu A_\mu
= \frac{1}{\sqrt{-g}} \partial_\mu \left( \sqrt{-g} g^{\mu\nu} A_\nu \right) 
= - \partial_t A_t + 3 H A_t + a^{-2} \sum_{i=1,2,3} \partial_i A_i \, .
\end{equation}
Then Eq.~(\ref{phtn1}) reduces to
\begin{equation}
\label{phtn4}
0 = \nabla^2 A^\nu + R^{\mu\nu} A_\mu\, .
\end{equation}
Furthermore because we are interested in the propagation of photon, we may assume, 
\begin{equation} 
\label{phtn5}
0 = A_t = \sum_{i=1,2,3} \partial_i A_i \, .
\end{equation}
Then the first eq.(\ref{phtn2}) and the gauge condition (\ref{phtn3}) are satisfied and 
the second equation in (\ref{phtn2}) gives 
\begin{equation} 
\label{phtn6}
0 = - \left(\partial_t^2 + H \partial_t \right) A_i + a^{-2} \bigtriangleup A_i \, .
\end{equation}

First, we consider the propagation in the de Sitter space-time, where the Hubble rate $H$ is a constant, 
$H=H_0$, and the scale factor is given by $a=\e^{H_0 t}$.
Because the propagation of the light is simpler than that of the gravitational wave, first we consider 
Eq.~(\ref{phtn6}).
We now assume the plane wave and separate the variable as follows, 
$A_i \propto \e^{i\bm{k} \cdot \bm{x}} \hat A_i(t)$ with the wave number $\bm{k}$, and replace 
$\bigtriangleup$ by $- k^2 \equiv - \bm{k}\cdot \bm{k}$.
Further we redefine a new variable $s$ by 
\begin{equation} 
\label{phtn7} 
s \equiv \e^{- H_0t} \, , 
\end{equation}
Eq.~(\ref{phtn6}) can be rewritten as
\begin{equation}
\label{phtn8}
0 = \left( \frac{d^2}{ds^2} + \frac{k^2}{H_0^2} \right) \hat A_i \, , 
\end{equation} 
whose solution is trivially given by 
\begin{equation} 
\label{phtn9} 
\hat A_i = A_{i0} \cos \left( \frac{k}{H_0} s + \theta_0 \right) \, .
\end{equation}
Here $A_{i0} $ and $\theta_0$ are constants.

Let us now consider the gravitational wave (\ref{LB3}) in the background of  de Sitter space-time.
We further define  variable $u$ and a function $l_{ij}$ by 
\begin{equation} 
\label{GWp1} 
u = \frac{k}{H_0}s \, ,\quad h_{ij}=s^{-\frac{1}{2}} l_{ij} \, .
\end{equation}
Then Eq.~(\ref{LB3}) can be rewritten as 
\begin{equation} 
\label{GWp2}
0 = \left( \frac{d^2}{du^2} + \frac{1}{u} + 1 - \frac{ \left(\frac{5}{2}\right)^2}{u^2}
\right) l_{ij} \, ,
\end{equation}
which is nothing but Bessel's differential equation, whose solution is given by the Bessel functions 
$J_{\pm \frac{5}{2}} (u)$.
As long as we consider the gravitational wave from the black hole merger 
\cite{Abbott:2016blz,Abbott:2016nmj} or neutron star \cite{TheLIGOScientific:2017qsa}, the magnitude 
of the variable $s$ in (\ref{phtn7}) is the order of unity.
On the other hand, the wave number $k$ should be much larger than the Hubble constant $H_0$, which 
is the present value of the Hubble rate $H$ and therefore $u$ in (\ref{GWp1} should be very large.
Using the asymptotic behavior of the Bessel function 
\begin{equation} 
\label{GWp3} 
J_\alpha (x) \to \sqrt{\frac{2}{\pi x}} \cos \left( x - \frac{2\alpha + 1}{4}\pi \right) \, , 
\end{equation} 
for large $x$, we find 
\begin{equation} 
\label{GWp4} 
h_{ij} \sim \frac{1}{s}  \cos \left( \frac{k}{H_0} s + \frac{\pm 5 +1}{4}\pi \right) \, .
\end{equation}
By comparing (\ref{GWp4}) with (\ref{phtn9}) for the propagation of the light, there is no difference 
between the propagating speed of the light and the gravitational wave.

We now consider the case that the scale factor $a(t)$  behaves as a power-law function, 
\begin{equation} 
\label{grv5}
a(t) = \left( \frac{t}{t_0} \right)^\alpha \, , 
\end{equation} 
with  $t_0$ and $\alpha$ real constants. Depending on the value of $\alpha$,
Eq.~(\ref{grv5}) involves a power law (superluminal) inflation  $(\alpha \geq 1)$, a Friedmannian 
(subluminal) evolution $(0<\alpha < 1)$, and a pole-like (phantom
\cite{Caldwell:1999ew,Caldwell:2003vq,Nojiri:2003vn}) behavior $(\alpha<0)$.
The universe described by (\ref{grv5}) can be realized by the scalar-tensor model in
(\ref{H7}) by substituting Eq.(\ref{grv5}) into the expressions for $\omega(\phi)$ and $V(\phi)$ 
(\ref{LB2}), that is 
\begin{equation} 
\label{LB2power} 
\omega (\phi)  = \frac{2\alpha }{\kappa^2 t_0^2 \phi^2} \, , \quad
V(\phi) = \frac{3\alpha^2 - \alpha}{\kappa^2 t_0^2 \phi^2} \, .
\end{equation}
In case that the expansion of the universe is generated by the perfect fluid with a constant equation of 
state parameter $w$, the power-law expansion  (\ref{grv5}) is also realized by 
\begin{equation} 
\label{GWp5} 
\alpha = \frac{2}{3\left( 1 + w \right)} \, .
\end{equation}
For the power-law case (\ref{grv5}), the Hubble rate $H$ and $\dot H$ are given by 
\begin{equation} 
\label{GWp6} 
H = \frac{\alpha}{t} \, , \quad \dot H = - \frac{\alpha}{t^2} \, .
\end{equation}
If we consider the gravitational wave from the black hole merger \cite{Abbott:2016blz,Abbott:2016nmj} 
or neutron star merger \cite{TheLIGOScientific:2017qsa}, we may consider $H$ to be a constant 
$H\sim H_0$.
Because Eq.~(\ref{GWp6}) shows $H^2 \sim \dot H$ as functions of $t$, one may take $\dot H$ to be 
a constant, $\dot H = H_1$,  as long as we can take $H^2$ to be a constant.
Note the term including $\dot H$ appears in Eq.~(\ref{LB3}) in addition to the term of $H^2$.
Then by regarding the terms $2\dot H + 6 H^2$ with constants, 
$2\dot H + 6 H^2 \sim 2 H_1 + 6 H_0^2$.
Then by using (\ref{GWp1}) with the variable $s$ in (\ref{phtn7}), which is defined by using the 
constant $H_0$, that is,  the Hubble constant in the present universe, instead of (\ref{GWp2}), we 
rewrite (\ref{LB3}), as follows, 
\begin{equation} 
\label{GWp2m}
0 = \left( \frac{d^2}{du^2} + \frac{1}{u} 
+ 1 - \frac{\left(\frac{5}{2}\right)^2 - \frac{2H_1}{H_0^2}}{u^2} \right) l_{ij} \, ,
\end{equation}
Then the solution is given by $J_{\pm \frac{5}{2} \sqrt{ 1 - \frac{4H_1}{25H_0^2}}}(u)$ and 
instead of (\ref{GWp4}), we find 
\begin{equation} 
\label{GWp4m} 
h_{ij} \sim \frac{1}{s}  \cos \left( \frac{k}{H_0} s + \frac{\pm 5\sqrt{ 1 + \beta}
+1}{4}\pi \right) \, , \quad \beta \equiv - \frac{4H_1}{25H_0^2}\, ,
\end{equation}
Hence, it does not appear the term including $\dot H$ in Eq.~(\ref{phtn6}), which describes the 
propagation of the light.
Thus, the ratio of the propagating light speed $c$  is identical with the propagating speed 
$c_\mathrm{GW}$ of the gravitational wave.
We should note that there appears the shift of the phase in (\ref{GWp4m}) compared with
(\ref{GWp4}) but it could be difficult to detect the phase.
Because
\begin{equation}
\label{GWpRR1}
\beta = - \frac{4}{25\alpha} = - \frac{6(1+w)}{25}\, , 
\end{equation} 
if one can find the phase by the observations, we may find the constraint on the equation of state 
parameter $w$ in the scalar-tensor theory.
In the radiation dominance epoch,  $w=\frac{1}{3}$ and in the matter dominance epoch, $w=0$. 
If current accelerating universe  corresponds to the asymptotic de Sitter space-time, $w$ should be $-1$.
 From Eq.~(\ref{GWpRR1}), we see that actually there is no difference of phase at the inflation and 
current universe but it may be some visible difference in radiation/matter dominated eras.

\section{$F(R)$ gravity  \label{SIV}}

Let us now investigate $F(R)$ gravity in the similar fashion.
Just for the simplicity, we assume that $F(R)$ is given by the power of the scalar curvature $R$, 
\begin{equation} 
\label{FR}
F(R) \sim R^m\, ,
\end{equation}
which gives the power-law scale factor (\ref{grv5}) \cite{Nojiri:2003ft}, where the exponent $\alpha$ 
is related with $m$ as follows, 
\begin{equation} 
\label{grv11} 
\alpha = - \frac{(m-1)(2m-1)}{m-2}\, .
\end{equation}
By using the conformal transformation
\begin{equation}
\label{grv12}
g_{\mu\nu} = \frac{1}{F'(R)} g_{\mathrm{E}\, \mu\nu}\, , 
\end{equation} 
the Jordan frame is mapped into the Einstein frame.
In \cite{Capozziello:2017vdi}, it has been shown that the gauge conditions
(\ref{H4}) do not
change as long as we consider the propagation of the gravitational wave.
For the power-law scale factor  (\ref{grv5}), the scalar curvature is proportional to $\frac{1}{t^2}$ and 
$F'(R)$ can be written as 
\begin{equation} 
\label{grv13}
F'(R) \sim F_0 \left( \frac{t}{t_0} \right)^{-2 (m - 1)}\, .
\end{equation}
Then the metric (\ref{grv12}) in the Einstein frame is given by 
\begin{equation} 
\label{grv14} 
ds^2_E = F_0 \left( \frac{t}{t_0} \right)^{-2 (m - 1)} \left[ - dt^2
+ \left( \frac{t}{t_0} \right)^{ - \frac{2(m-1)(2m-1)}{m-2}}
\sum_{i=1,2,3} \left( dx^i \right)^2 \right]\, .
\end{equation}
One may define the time coordinate $t_E$ in the Einstein frame as follows, 
\begin{equation} 
\label{grv15} 
t_E = t_{E0}\left( \frac{t}{t_0} \right)^{ 2- m } \, .
\end{equation}
Here $t_{E0}$ is a constant defined by
\begin{equation}
\label{grv16}
t_{E0} \equiv \frac{ \sqrt{F_0 t_0 }}{ 2 - m} \, .
\end{equation}
By using $t_E$, the metric in the Einstein frame can be rewritten as 
\begin{equation} 
\label{grv17} 
ds^2_E = - dt_E^2
+ \left( \frac{t_E}{t_{E0}}\right)^{6 \frac{(m-1)^2}{(m-2)^2}}
\sum_{i=1,2,3} \left( dx^i \right)^2 \, .
\end{equation}
Then $\tilde\alpha \equiv 3 \frac{(m-1)^2}{(m-2)^2}$ can be identified with $\alpha$ in (\ref{grv5}) 
in the Einstein frame, what is different from $\alpha$ (\ref{grv11})
in the Jordan frame.

The speed of  the gravitational wave should not be changed when transition to different frame is made.
The exponent $\tilde \alpha$ is related with $w$ by (\ref{GWp5}), the equation of state parameter 
$\tilde w$ in the Einstein frame is given by 
\begin{equation} 
\label{FGWpr1} 
\tilde w +1 = \frac{2}{3\tilde\alpha} = \frac{2(m-2)^2}{9(m-1)^2}\, .
\end{equation}
Then the parameter $\beta$ (\ref{GWp4m}) in the phase is now given by 
\begin{equation} 
\label{GWpRR2} 
\beta = - \frac{4}{25\tilde \alpha} = \frac{12 (m-2)^2}{25(m-1)^2}\, , 
\end{equation} 
If $w\sim -1$ as in the the Planck satellite data \cite{Ade:2015xua}, 
\begin{equation} 
\label{GWp11} 
w=-1.019^{+0.075}_{-0.080} \, , 
\end{equation} 
we find $m\sim 2$ and $w+1\sim \frac{2(2-m)}{9}$, and therefore 
\begin{equation} 
\label{GWpRR3} 
\beta \sim \frac{243(w+1)^2}{25}\, , 
\end{equation} 
which is rather different from the Eq.(\ref{GWpRR1}).
Therefore even if $w$ is the same, there appears the difference in the phase.
If one can detect the gravitational wave phase, we can distinguish the $F(R)$ gravity from 
the scalar-tensor theory.
It also follows from (\ref{GWpRR3}) that the phase vanishes for the exact de Sitter space-time but even 
for the asymptotically de Sitter space-time, the phase does not vanish although it is, of course, very 
small.

\section{Summary}

In this paper, we have investigated the propagation of the gravitational wave in the scalar-tensor theory 
and $F(R)$ gravity.
Although we do not include the derivative coupling with the curvature in the scalar-tensor theory, we 
have shown that the propagation of the gravitational wave depends on the modified gravity 
as the scalar-tensor theory and $F(R)$ gravity theory. 
Some difference in the gravitational 
wave propagation phase if compare with the light one is observed.

It could be interesting to apply the above formulation to other modified gravities like $f(\mathcal{G})$ 
gravity whose action is given by 
\begin{equation} 
\label{GB1b} 
S=\int d^4x\sqrt{-g} \left(\frac{1}{2\kappa^2}R
+ f(\mathcal{G}) + \mathcal{L}_\mathrm{matter}\right)\, .
\end{equation}
Here $f(\mathcal{G})$ is a function of the Gauss-Bonnet invariant $\mathcal{G}$, which is defined by 
\begin{equation} 
\label{GB}
\mathcal{G}=R^2 -4 R_{\mu\nu}
R^{\mu\nu} + R_{\mu\nu\xi\sigma} R^{\mu\nu\xi\sigma}\, .
\end{equation}
As  the Gauss-Bonnet invariant $\mathcal{G}$ is given by the squares of the curvatures, the correction 
to the propagation of the gravitational wave could be small.
The question is if the gravitational wave speed can exceed the light speed in the $f(\mathcal{G})$ 
gravity model or only change of gravitational wave propagation phase occurs. This will be studied 
elsewhere.

In the case of the real matter and radiation, we know the operators corresponding to 
$\delta T_{\mu\nu}$ in (\ref{H6}) because  the Lagrangian densities for the matter and the radiation 
are known.
Then by using the statistical physics etc, we may estimate the average of the operators.
The obtained expression for $\delta T_{\mu\nu}$ could be different from those in the scalar-tensor 
theory or $F(R)$ gravity theory.

Finally we consider the possibility to detect the shift of the phase which we found in this paper 
by cosmological observations.
One way could be to observe the difference between the phases of two kinds of gravitational wave:
one gravitational wave arrives at the earth directly from the source and other gravitational wave 
goes through nearby galaxies or any massive object before arriving at the earth.
If the EoS parameter $w$ is in the range $-1<w<0$, the density of the dark energy decreases near 
the galaxy or any massive object due to negative pressure. Therefore there occurs the difference 
in the phases for these two kinds of the gravitational waves.
As an example, we may consider the density of the dark energy around the spherically symmetric 
massive object and we may assume the Schwarzschild type metric, 
\begin{equation} 
\label{Sch1}
ds^2 = - \e^{2\nu(r)} dt^2 + \e^{-2\nu(r)} dr^2 + r^2 d\Omega^2\, .
\end{equation}
Here $d\Omega^2$ expresses the metric of the two-dimensional unit sphere.
By assuming the energy density $\rho$ and the pressure $p$ of the dark energy only depend on 
the radial coordinate $r$, the $r$ component of the conservation law $\nabla^\mu T_{\mu r}=0$ 
for the energy-momentum tensor gives, 
\begin{equation} 
\label{Sch2} 
\frac{d p}{dr} + \nu' \left( p + \rho \right) = 0 \, .
\end{equation}
Then if $p=w\rho$, we find
\begin{equation}
\label{Sch3}
\rho=\rho_0 \e^{- \left( 1 + \frac{1}{w} \right) \nu }\, .
\end{equation}
In the case of the Schwarzschild metric $\e^{2\nu} = 1 - \frac{2M}{r}$, $\e^{2\nu}$ is 
a monotonically increasing function with respect to $r$ outside of the horizon $r>2M$. 
Then Eq.~(\ref{Sch3}) tells that if $-1<w<0$, $\rho$ is an increasing function of $r$, that is, $\rho$ 
becomes smaller near the massive object. 
In case of the cosmological constant where $w=-1$, 
there does not occur the change of the density, which is clear from Eq.~(\ref{Sch3}). 
In order to find the shift of the phase, we may observe the interference in the gravitational waves. 
For example, we may consider the case that there is a massive object between the earth and 
the source of the gravitational wave. Then there occurs the interference between the gravitational wave 
which passed near the massive object and the wave which passed the region far 
from the massive object. 
This interference changes the waveform in addition to the possible gravitational lensing effect.  
When we obtain Eq.~(\ref{Sch3}), we have assumed that the dark energy is given by the perfect 
fluid whose EoS parameter is a constant $w$.
The assumption could be changed if we consider the scalar-tensor theory or $F(R)$ gravity 
as the dark energy, which could be also an interesting future work.

\section*{Acknowledgments.}

This work is supported (in part) by
MEXT KAKENHI Grant-in-Aid for Scientific Research on Innovative Areas ``Cosmic Acceleration''  
(No. 15H05890) (SN), by MINECO (Spain), project FIS2016-76363-P and JSPS S17116 
short-term fellowship (SDO) and by CSIC I-LINK1019 Project(SN and SDO). 
This research was started while SN was visiting the Institute for Space Sciences, ICE-CSIC 
in Bellaterra, Spain (SN thanks Emilio Elizalde and SDO, and the rest of the members of the ICE 
for very kind hospitality).


\end{document}